\documentstyle[12pt,epsfig]{article}

\def\d{\delta}
\def\n{\nu}
\def\be{\begin{equation}}
\def\ee{\end{equation}}
\def\bea{\begin{eqnarray}}
\def\eea{\end{eqnarray}}
\def\n{p(S)}
\def\m{\m_\nu}

\author{F. Y. Wu,$^1$ C. King,$^2$ and W. T. Lu$^1$ \\
  \\
$^1$Department of Physics\\
$^2$Department of Mathematics\\
  Northeastern University \\
 Boston,
Massachusetts 02115}

\title{On the rooted Tutte
 polynomial}

\begin{document}
\maketitle

\bigskip
\begin{abstract}
 The Tutte  polynomial
 is a generalization of   the chromatic  polynomial of  graph colorings.
Here we present an extension called the
  rooted Tutte polynomial, which is defined on a graph
where one or more
 vertices  are   colored with prescribed  colors.
We establish a number of results pertaining to the rooted Tutte polynomial,
including
 a duality relation
in the case that  all  roots reside
around a single face of a planar graph.
The connection with the Potts model is also reviewed.

\end{abstract}
\vskip 1cm
\noindent

\newpage

\section{The  Tutte polynomial}
Consider a finite graph $G$ with  vertex set $V$ and  edge set $E$.
A spanning subgraph  $G'(S)\subseteq G$ is a subgraph of $G$ containing all
members of
$V$ and an edge set $S\subseteq E$.
Let $C$ be a set of $q$ distinct colors.  A $q$-coloring of $G$
is a coloring of the vertices in
$V$ such that two vertices connected by an edge bear different
colors.  It is well-known that
the number of $q$-colorings of $G$ is given by the chromatic polynomial
\cite{birkhoff}
\be
P(G; q) =  \sum_{S\subseteq E} q^{\n} (-1)^{|S|}, \label{ch}
\ee
 where  $\n$ is the number of
components in the spanning subgraph $G'(S)$.
Alternately, we can regard
(\ref{ch}) as generating  colorings of components of spanning
subgraphs of $G$ with $q$ colors with an edge weight $-1$.

As an extension of the chromatic polynomial,
 Tutte \cite{tutte54,tutte67,tutte84}
introduced  what is now known as the Tutte polynomial
\be
Q(G;t,v)= \sum_{S \subseteq E} t^{\n} v^{|S|-|V|+\n}.\label{tutte}
\ee
 Indeed, one has the relation
 \bea
 P(G; q)=(-1)^{|V|}Q(G;-q,-1)  . \label{chromatic}
\eea
In view of  (\ref{chromatic}), it is  useful to write (\ref{tutte}) as
\be
Q(G;t,v)= v^{-|V|}\sum_{S \subseteq E} (vt)^{\n} v^{|S|},\label{tutte1}
\ee
so that for $vt=q=$ positive  integers,  the Tutte polynomial (\ref{tutte1})
 generates
  colorings of  components of
spanning subgraphs of $G$ with $q$ colors and edge weights
 $v$, instead of $v=-1$.

For planar $G$ with dual graph $G_D$, it is well-known that
the Tutte polynomial possesses the duality relation
\be
v \ Q(G;t,v) =
t \ Q(G_D;v,t),  \label{dual}
\ee
 a relation first observed by
Whitney  \cite{whitney}.

\section{The rooted Tutte polynomial}
We extend the definition (\ref{tutte1}) to  a rooted Tutte polynomial.

A vertex is rooted, or is a root,
 if it is  colored with a prescribed (fixed) color.
A graph is rooted if it contains rooted vertices.
   Let $R$ denote a set of $n$ roots located at vertices  $\{r_1, r_2,
\cdots,r_n\}$.
A {\it color configuration} is a map
$x:R\longmapsto C$, and as a convenient shorthand we write $x(r_{i})=x_{i}$
for $i=1,2,\cdots,n$.
 A component of a spanning subgraph  is {\it exterior}
if it  contains one or more roots, and is {\it interior} otherwise.
An exterior component is {\it proper} if all roots in the component
are of the same color. A spanning subgraph $G'(S)$ is proper if all its
exterior components are proper.
An edge set $S_x\subseteq E$ is proper
if   the spanning subgraph $G'(S_x)$ it generates is proper.

 For a prescribed color
configuration $\{x_1,x_2,\cdots,x_n\}$ of the $n$ roots,
we introduce in analogy to (\ref{tutte1})   the  {\it rooted} Tutte
 polynomial\footnote{Strictly speaking, it is the expression
$v^{|V|}Q_{x_1x_2\cdots x_n}
(G;t,v)$ which is a polynomial in $v$ and $t$.}
   \be
Q_{x_1x_2\cdots x_n}
(G;t,v)= v^{-|V|}\sum_{S_x \subseteq E}
(vt)^{p_{\rm in}(S_x)} v^{|S_x|},\label{tutte2}
\ee
where the summation is taken over  all proper edge sets $S_x$, and
 $p_{\rm in}(S_x)$ is the number of interior components of $G'(S_x)$.
  Thus, as in (\ref{chromatic}), we have for positive integral $q$ the
relation
\bea
(-1)^{|V|} Q_{x_1x_2\cdots x_n}
(G;-q,-1) &=& {\rm the\>\>number\>\>of\>\>}q{\rm -colorings\>\>of\>\>}G
{\rm \>\>with}
\nonumber \\
&& {\rm color\>\>configuration\>\>}\{x_1,x_2,\cdots, x_n\}.
\eea
Clearly, the  expression (\ref{tutte2}) depends  on how the $n$
roots are partitioned into subsets of different colors, and the actual colors
do not enter the picture.

The coloring configuration $\{x_1, x_2, \cdots, x_n\}$ induces a partition
 $X$  of $R$
into blocks (subsets) such that all roots in one block are
of one color, and colors of different blocks are different.
Namely, two elements $r_{i},r_{j}\in R$ belong to the same block of $X$ if
and only if they have the same prescribed color $x_i=x_j$.
 Consider now the summation in (\ref{tutte2}).
Let $G'(S)$ be any (not necessarily
proper) spanning subgraph of $G$. The connected components of
$G'(S)$ induce  a partition on the set of vertices
$V$ of $G$. We get hence also a partition $\pi(S)$ on the
set of rooted vertices $R$
by restricting this partition
to $R$. Clearly, the spanning subgraph $G'(S_x)$ is proper if and only if
the partition $\pi(S_x)$ is a refinement of the partition $X$.
 It follow that we can rewrite (\ref{tutte2}) as
\be
Q_{X}(G;t,v) = \sum _{X'\preceq X} F_{X'}(G;t,v), \label{qq}
\ee
where
\be
F_{X'}(G;t,v)\equiv
v^{-\vert V\vert}\sum_{S_x\subseteq E,\ \pi(S_x)=X'}
(vt)^{p_{\rm in}(S_x)}v^{\vert S_x\vert}.  \label{qqq}
\ee
Here, we have  abbreviated
$Q_{x_1 x_2 \cdots x_n}$  by $Q_X$,
which is permitted since  the actual colors do not
  enter the picture at this point. Also it is understood that $G$ is now
a rooted graph, with root set $R$.

The expression (\ref{qq})  assumes the  form
  of a transformation of a partially ordered
set.  Its  inverse   is given by  the M\"obius inversion
\be
F_X(G;t,v) = \sum _{X'} \mu(X',X)Q_{X'}(G;t,v) , \label{inverse}
\ee
where  \cite{lint}
\bea
 \mu(X',X) &=& (-1)^{|X'|-|X|} \prod_{{\rm blocks \>\>}\epsilon \> X}
     (n_{b}(X')-1)!, \hskip 1cm {\rm if}\>\> X'\preceq X \nonumber \\
  &=& 0, \hskip 5.9cm {\rm otherwise,}
\label{mu}
\eea
 $n_{b}(X')$ being the number of blocks of $X'$ that are contained in
the block $b$ of $X$.
Note that for $n=1$ we have $|X|=|X'|=1$,    $p_{\rm in}(S_x) =p(S_x) -1$,
and all edge sets $S\subseteq E$ are proper.  Hence we have
 \be
F_X(G;t,v) = Q_X(G;t,v) = (vt)^{-1}  Q(G;t,v), \hskip 1cm n=1. \label{n1}
\ee
This completes the definition  and general description
of the rooted Tutte polynomial for any graph $G$.

\section{Planar graphs}
From here on we
consider $G$ being planar  with the $n$ roots residing around a single
face of $G$.  Without the loss of generality, we can
choose the face to be the infinite face and order the roots in the
sequence $\{r_1,r_2,\cdots,r_n,r_1\}$ as shown in Fig. 1.
  A partition $X$ of the $n$ roots is
{\it non-planar} if two roots  of one block separate two roots of another
block in the cyclic sequence.
Otherwise $X$ is {\it planar}.
For a given $n$, there are $b_n$ partitions, where  \cite{wu2}
\be
b_n = \sum_{m_\nu =0}^\infty \biggl[ n!\big/\prod_{\nu=1}^{\infty} (\nu
!)^{m_\nu }
m_\nu !\biggr], \hskip 1cm \sum_{\nu=1}^\infty \nu m_\nu =n,
\ee
and of the $b_n$ partitions
\be
c_n={{(2n)!}/ {n! (n+1)!}}
\ee
 are planar \cite{templieb,tutte93}.
We shall adopt the convention of writing $X=\{ij,k\ell\cdot \cdot,\cdots\}$
for colors $\{x_i=x_j, \ x_k=x_\ell=\cdot\cdot,\cdots\}$, with
$\{ij\}, \{k\ell \cdot\cdot\}, \cdots$
each in order
\cite{tutte93}.  For example, two partitions for $n=5$ are
 \bea
&&X_1=\{123, 4, 5\}, \hskip 1cm |X_1|=3,\hskip 1cm {\rm planar},\nonumber \\
&& X_2= \{24, 351\}, \hskip 1.2cm |X_2|=2, \hskip 1cm {\rm non-planar}.
\label{partexam}
\eea
Now if
 $G$ is planar and
$X'$ is non-planar then  by definition the summand in (\ref{qqq}) is empty and
one has $F_{X'}(G;t,v)=0$.
Thus we have

\smallskip
\noindent
{\it Proposition 1:}

For planar $G$
\be
F_X(G; t,v) =0, \hskip 1cm {\rm if\>\>}X{\rm \>\>is \>\>
non-planar}.\label{proposition}
\ee
This proposition was first 
  established  in \cite{wu2}
for the  Potts model correlation function (see section 6) by considering 
its graphical expansion similar to the consideration given in the above.
 As a consequence of  Proposition 1 and the use of (\ref{inverse}), we now have

\noindent
{\it Corollary 1}:

Rooted Tutte polynomials associated with non-planar partitions can be 
written as linear combinations  of the rooted Tutte 
polynomials associated with (refined) planar partitions.

\begin{figure}[htbp]
\center{\rule{5cm}{0.mm}}
\rule{5cm}{0.mm}
\vskip -1.2cm
\hskip 0.5cm
\epsfig{figure=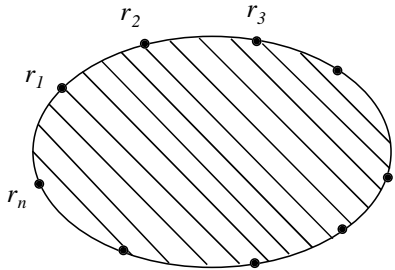,height=6.5in}
\vskip -10.8cm
\caption{A planar graph $G$ with $n$ roots.  The graph is denoted by
the shaded region and  the   $n$ roots by
 the black circles.
\label{fig:fig1}}
\end{figure}
\vspace*{-2pt}

Corollary 1 leads to the sum-rule identities reported in \cite{wu2} for
the Potts correlation function.  
In the case of $n=4$, for example, the identity 

\noindent
$F_{\{13,24\}}(G;t,v)=0$  leads to  the
 sum rule \cite{wu2}
\be
Q_{\{13,24\}}(G;t,v) = Q_{\{13,2,4\}}(G;t,v)+Q_{\{1,3,24\}}(G;t,v)
-Q_{\{1,2,3,4\}}(G;t,v).
\ee
  From here on we shall restrict our considerations to 
rooted Tutte polynomials associated with the $c_n$ planar partitions only.

\section{The graph $G^*$}
The rooted Tutte polynomial (\ref{tutte2}) possesses a duality relation
for planar graphs,
which
 relates the rooted Tutte polynomial on a  graph
$G$ to that of a related graph  $G^*$.
Here we define $G^*$.
Starting from a planar $G$, place an extra vertex $f$ in the infinite
face and connect it to each root of $G$ by an edge.
This gives a new graph $G''$, which has one more vertex
 than $G$ and $n$ additional edges.
The dual graph of $G''$ is also planar, and it has a face $F$ containing the
extra vertex $f$.  Now remove the $n$ edges on the boundary of $F$, and the
resulting graph is $G^*$.

It is readily seen that the graph $G^*$ has
\be
|V^*|=|V_D|+n-1  \label{vstar}
\ee
vertices where $|V_D|$ is the number of vertices of $G_D$, the dual of $G$,
and there is  a one-one correspondence between the  edges of $G$ and
$G^*$.
We denote the set of $n$ vertices $\{r^*_1,r^*_2,\cdots,r^*_n\}$
of $G^*$ surrounding the face $F$
  by $R^*$, with  $r^*_i$ residing
 between the two edges
$\langle f, r_{i-1} \rangle$ and $\langle f, r_{i} \rangle$ of $G''$, where
$r_0=r_n$.
An example of a $G$ and the
related $G^*$ for $n=4$ is shown in Fig. 2.
Clearly, the relation of  $G$ to $G^*$ is reciprocal,
namely,  we have $(G^*)^*=G$.

\begin{figure}[htbp]
\center{\rule{5cm}{0.mm}}
\rule{5cm}{0.mm}
\vskip -1.2cm
\hskip 0.5cm
\epsfig{figure=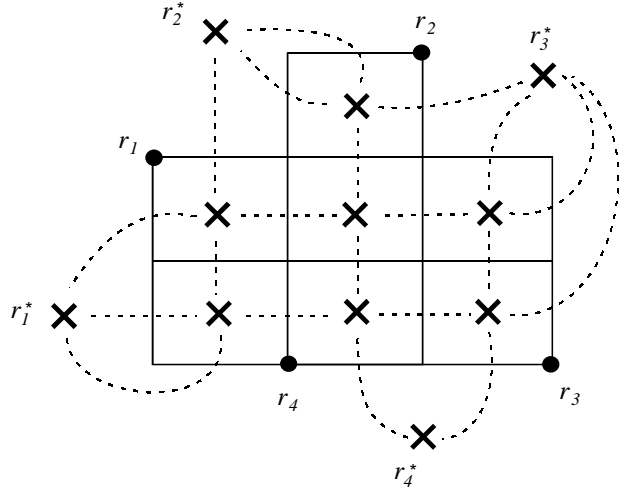,height=6.5in}
\vskip -8.5cm
\caption{A graph $G$ (solid lines) and the related
graph $G^*$ (broken lines) for   $n=4$.  Black circles denote
roots of $G$ and crosses denote vertices of $G^*$.
\label{fig:fig2}}
\end{figure}
\vspace*{-2pt}

 Now each planar partition $X$ of $R$ induces a
 partition $X^*$ of the set $R^*$   \cite{luwu}.
In order to define $X^*$, for each block $b$ of $X$ we
choose a point in the infinite face of $G$, and connect all roots $r_i$ in
$b$ to this point by drawing new edges. Because $X$ is planar, the points
for the blocks can be chosen so that the edges of different blocks do not
cross, and the resulting extended graph is still planar. So this process
divides the infinite face into regions.
The induced partition $X^*$  is then described by the condition that
  all roots   of $R^*$ in one region are regarded as belonging to one block
of $X^*$.  Alternately, for another way
of defining $X^*$, 
 let $P_{ij}$, $1 \leq i < j \leq n$, be a partition of $R$ into
two blocks,  the sets $\{r_{i}, \cdots, r_{j-1}\}$ and
$\{r_{j}, \cdots, r_{i-1}\}$,
where all numbers are
modulo $n$.  Then, the partition $X^*$ induced by $X$ is defined by the
condition that the two roots
$ r^*_{i}$ and $ r^*_{j}$
belong to the same block in $X^*$ if and only if $X$ is a refinement of
$P_{ij}$.

 We write $X\to X^*$ if $X$ induces $X^*$.
 Clearly,  $X^*$ is planar,  and we have
\be
|X| + |X^*| =n+1. \label{xy}
\ee
     For  the  planar partition $X_1$ in (\ref{partexam}), for example,
we have
\be
X_1=\{123,4,5\}\to X^*_1=\{2,3,451\}, \hskip 1cm |X_1|= |X^*_1|= 3.
\ee
However, as a result of our labelling convention, the
color configuration of the
partition
$(X^*)^*$ further induced by $X^*$
is a cyclic shift of that of $X$, namely,
\be
\{x_1,x_2,\cdots,x_n\} \to X^* \to \{x_n,x_1,\cdots,
x_{n-1}\} . \label{shift}
\ee
In the example above, for instance, we have
\be
\{123,4,5\}\to \{2,3,451\}\to \{234,5,1\}.
 \ee

Finally, there is a one-to-one correspondence between the
edge sets $E$ of $G$ and
$E^*$ of $G^*$,
an edge set $S\subseteq E$ defines a ``complement" edge set $S^*\subseteq E^*$
by the condition that
an edge is included in $S^*$ if and only if its corresponding edge is not
included
in $S$.  Clearly, we have $(S^*)^*=S$.

\section{The duality relation}
The rooted Tutte polynomial arises in statistical physics
as the correlation function of the Potts model (see next section).
In a recent paper \cite{luwu} we
have established a duality relation  for
the Potts correlation function for planar $G$.
  However,   the proof of the duality relation given in \cite{luwu}
is cumbersome and not easily deciphered in graphical terms.
  Here we re-state the results as two propositions in the
context of the rooted Tutte polynomial, and present
direct graph-theoretical proofs of the propositions.

 \smallskip
\noindent
{\it Proposition 2:}

For planar $G$ and $G^*$ and the associated planar partitions $X\to X^*$,  we have
\be
 v^{|X|}F_X(G;t,v)= t^{|X^*|}F_{X^*}(G^*;v,t).  \label{propo2}
\ee

\noindent
{\it Proof}:
Let $S_x$ be a  proper edge set on $G$.
We have the Euler relation
\be
|S_x|+  |S_x^*|= |V|+ |V_D|  -2 \label{euler}
\ee
and, after eliminating $n$ and $|V_D|$ using (\ref{vstar}), (\ref{xy}) and
(\ref{euler}), the identity
\be
|S_x| +|X| -|V| = |S_x^*| +|X^*| -|V^*|, \label{iden}
\ee
which holds for any  proper edge set $S_x$.
Note that we have also the fact
\be
\pi(S_x) = X \hskip .7cm {\rm if\>\>and\>\>only\>\>if}
\hskip .7cm   \pi(S^*_x) = X^*.\label{pp}
\ee

Let $c(S^*_x)$  the number of independent circuits in the spanning 
subgraph $G'(S_x^*)$.  Then we have 
\be
p (S_x) = c(S^*_x) + |X|. \label{i1}
\ee
Also, starting from the $|V^*|$ isolated vertices on $G^*$, one
constructs $G'(S^*_x)$
by drawing edges
 of $S^*_x$ on $G^*$ one at a time.  Since each edge reduces
the number of components by one except when the adding of an edge
 completes  an independent circuit,
one has also
\be
p(S_x^*) = |V^*| - |S^*_x| + c(S^*_x). \label{i2}
\ee
 Eliminating
$c(S^*_x)$ using (\ref{i1})  and (\ref{i2})
and making use of the relations
\bea
p(S_x) &=& p_{\rm in}(S_x) + |X|, \nonumber \\
p(S_x^*) &=& p_{\rm in}(S_x^*) + |X^*|, \label{i4}
\eea
one obtains 
\be
p_{\rm in}(S^*_x) = p_{\rm in}(S_x) +|V^*|-|S_x^*|  -|X^*|.\label{i5}
\ee
The Proposition 2 now follows from the substitution of (\ref{i5})
into the right-hand side of (\ref{propo2}) where, explicitly,
\be
F_{X^*}(G^*;v,t)=
t^{-\vert V^*\vert}\sum_{S_x^*\subseteq E^*,\ \pi(S_x^*)=X^*}
(vt)^{p_{\rm in}(S^*_x)}t^{\vert S^*_x\vert},  \label{qqq1}
\ee
   and the use
of the identities  (\ref{iden}) and (\ref{pp}).
 This completes the proof of Proposition 2.

Proposition 2 was first conjectured in \cite{wu2}
 and established later in \cite{luwu} in the context of Potts
correlation functions (see next section)
without the explicit reference to the polynomial form (\ref{qqq}).

\smallskip
{\it Remark}:
 For $n=1$,  the duality relation (\ref{propo2})
for the rooted Tutte polynomial becomes the duality relation (\ref{dual}) for
the  Tutte polynomial.  This is a consequence of  (\ref{n1}).

\smallskip
\noindent
{\it Proposition 3}:

1. The rooted Tutte polynomials associated with the $c_n$ planar
partitions for $G$ and  $G^*$
are related by the  duality transformation
\be
Q_X(G;t,v) = \sum_Y {\bf T}_n(X,Y) Q_Y(G^*;v,t),\label{propo3}
\ee
where ${\bf T}_n$ is a $c_n\times c_n$ matrix with elements
\be
{\bf T}_n(X,Y)= t^{n+1}
 \sum _{X'\preceq X} (vt)^{-|X'|} \mu(Y,Y'),\hskip 1cm X' \to Y'.
\ee

2. The matrix ${\bf T}_n$ satisfies
 the identity
\be
[{\bf T}_n]^2 (X,X') = \d(x_1, x_2')\d(x_2, x_3')
\cdots \d(x_n, x_1').  \label{shift1}
\ee

\noindent
{\it Proof}:
The transformation (\ref{propo3}) follows
by combining
(\ref{qq}) and (\ref{inverse})
with Proposition 2,
and its uniqueness is ensured by the uniqueness of
the M\"obius inversion.
The property (\ref{shift1}) is a consequence
of (\ref{shift}).

Proposition 3 was first given in 
  \cite{luwu} in the context of the Potts correlation function (see next section).
Explicit expression of {\bf T}$_n$ for
$n=2,3,4$ can be found in \cite{luwu} and \cite{wu1}.

\section{The Potts and the random cluster models}
It is well-known in statistical physics that the Tutte polynomial gives
rise to the
partition function of the Potts model
\cite{wu}.  In view of the prominent role played by the Potts model
in many fields in physics, it is useful to review this
equivalence and
  the further equivalence of the rooted Tutte polynomial with the
Potts correlation  function.

 The   $q$-state
Potts model \cite{potts} is a spin model defined on a graph $G$.
The spin model consists of
 $|V|$  spins placed at the vertices of $G$ with  each spin
taking on $q$ different
states and interacting with spins connected by edges.
 Without going into details of the physics \cite{wu} which
lead to the Potts model, it suffices for our purposes to define
the Potts partition function
 \be
 Z(G;q,v) \equiv  \sum_{S \subseteq E} q^{\n} v^{|S|}, \label{pottsz}
\ee
the  $n$-point partial partition function
  \be
Z_{X}
(G;{q},v)\equiv \sum_{S_x \subseteq E} q^{p_{\rm in}(S_x)}
v^{|S_x|},\label{pottsp}
 \ee
and the $n$-point correlation function
\be
P_n(G;x_1,x_2,\cdots,x_n) = P_n(G;X)\equiv Z_{X}
(G;{q},v)\big/Z(G;q,v), \label{pottsc}
 \ee
where again, in analogy to notation
in Sections 1 and 2, we have denoted the color configuration
$\{x_1,x_2,\cdots, x_n\}$
by the associated partition $X$.
More generally, for any real or complex $q$, the partition function
(\ref{pottsz})
defines the random cluster model of Fortuin and Kasteleyn \cite{fk},
which coincides with the  Potts model for integral $q$.

Relating this to the Tutte polynomial, we  now have
\bea
 Z(G;q,v) &=&    v^{|V|} Q(G;t,v)  \nonumber  \\
 Z_X(G;q,v) &=& v^{|V|} Q_{X}(G;t,v) \nonumber \\
P_n(G;X)&=&
  Q_{X}(G;t,v)
\big/ Q (G;t,v),
  \label{correlation}
\eea
for $q=vt$.
  The duality relation (\ref{dual}) for the Tutte polynomial then
implies the following duality relation for the Potts partition function
 \cite{luwu,potts,wuwang}
  \be
v^{1-|V|} Z(G;q,v) = (v^*)^{1-|V_D|} Z(G_D;q,v^*), \label{dualpotts}
\ee
where
\be
vv^*=q.
\ee
 One further defines  the dual correlation function
\be
P^*_n(G^*;X^*) \equiv q\ Z_{X^*}(G^*;q,v^*)
\big/ Z(G_D; q,v^*),
\ee
and also the functions $A_{X}$ and $B_{X^*}$ by
  \be
P_n(G;X) = \sum_{X'\preceq X} A_{X'}(G;q,v)
\ee
and
\be
P^*_n(G^*;X^*) = \sum_{{X^*}'\preceq X^*}B_{{X^*}'}(G^*;q,v^*).
\ee
  Then, Proposition 2 leads to   the relation
\be
A_X(G;q,v)=q^{-|X|}B_{X^*}(G^*;q,v^*),\hskip 1cm X\to X^*.
\ee
which is the main result of \cite{luwu}.

\section{Summary and discussions}
We have introduced the rooted Tutte polynomial (\ref{tutte2}) as a two-variable
 polynomial  associated with a rooted graph
and deduced a number of pertinent results.

Our first result  is that the rooted Tutte polynomial
assumes  the form  (\ref{qq}) of a partially order set for which
the inverse can be uniquely determined.
For planar graphs and all roots residing
surrounding  a single face, we showed that (Proposition 1)
the inverse function vanishes for
non-planar partitions of the roots.  We further showed that
the inverse function
 satisfies the duality
relation (\ref{propo2}) (Proposition 2) which, in turn,  leads  to
the duality  (\ref{propo3}) for the rooted
Tutte polynomial (Proposition 3). We also reviewed the connection
of the Tutte and rooted Tutte polynomials with the Potts model
in statistical physics.

Finally,
we remark that  results reported here have previously been
obtained in \cite{wu2} and \cite{luwu} in the context of the Potts correlation function.
Here, the results are reformulated as properties of
the rooted Tutte polynomial and thereby
permitting graph-theoretical
 proofs.

 \vskip 8 mm
\centerline{ACKNOWLEDGMENTS}

\medskip
We are  grateful to the referee for providing an independent proof of
Proposition 1 and  numerous suggested  improvements on an earlier version of
this paper.
  This work is supported in part
by the National Science Foundation grants
DMR-9614170 (FYW and WTL) and DMS-9705779 (CK).

\end{document}